\documentclass[twocolumn,floatfix,superscriptaddress,showpacs]{revtex4}
\usepackage{graphicx}
\usepackage{amsmath}
\usepackage{bm}
\begin{document}

\title{
Correspondence between Andreev reflection and Klein tunneling in bipolar graphene}
\author{C. W. J. Beenakker}
\affiliation{Instituut-Lorentz, Universiteit Leiden, P.O. Box 9506, 2300 RA Leiden, The Netherlands}
\author{A. R. Akhmerov}
\affiliation{Instituut-Lorentz, Universiteit Leiden, P.O. Box 9506, 2300 RA Leiden, The Netherlands}
\author{P. Recher}
\affiliation{Instituut-Lorentz, Universiteit Leiden, P.O. Box 9506, 2300 RA Leiden, The Netherlands}
\author{J. Tworzyd{\l}o}
\affiliation{Institute of Theoretical Physics, Warsaw University, Ho\.{z}a 69, 00--681 Warsaw, Poland}
\date{November 2007}
\begin{abstract}
Andreev reflection at a superconductor and Klein tunneling through an \textit{n-p} junction in graphene are two processes that couple electrons to holes --- the former through the superconducting pair potential $\Delta$ and the latter through the electrostatic potential $U$. We derive that the energy spectra in the two systems are identical, at low energies $\varepsilon\ll\Delta$ and for an antisymmetric potential profile $U(-x,y)=-U(x,y)$. This correspondence implies that bipolar junctions in graphene may have zero density of states at the Fermi level and carry a current in equilibrium, analogously to superconducting Josephson junctions. It also implies that nonelectronic systems with the same band structure as graphene, such as honeycomb-lattice photonic crystals, can exhibit pseudo-superconducting behavior.  
\end{abstract}
\maketitle

\begin{figure}[tb]
\includegraphics[width=0.7\linewidth]{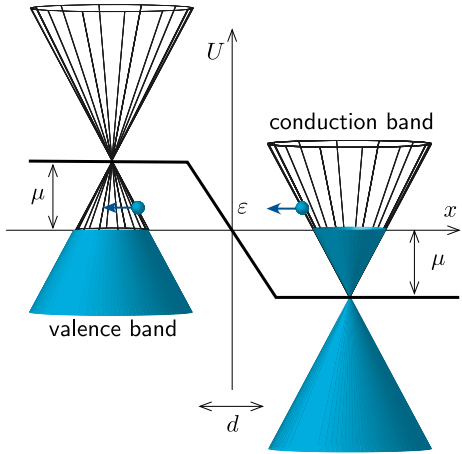}
\caption{\label{Klein_tunneling}
Conical band structure in graphene at two sides of a potential step (height $2\mu$, width $d$), forming an \textit{n-p} junction. In equilibrium, all states below the Fermi level (indicated in blue) are filled and all states above are empty. Klein tunneling is the interband tunneling of an electron from the conduction band in the \textit{n} region (blue ball at the right) into the valence band of the \textit{p} region (blue ball at the left). In this work we show that the low-energy excitation spectrum of a symmetric \textit{n-p} junction is the same as that of an NS junction, obtained by replacing the region $x<0$ by a superconductor.
}
\end{figure}

Tunneling through an \textit{n-p} junction in graphene is called Klein tunneling \cite{Kat06,Che06,Che07} with reference to relativistic quantum mechanics, where it represents the tunneling of a particle into the Dirac sea of antiparticles \cite{Kle29}. Klein tunneling in graphene (see Fig.\ \ref{Klein_tunneling})  is the tunneling of an electron from the conduction band into hole states from the valence band --- which plays the role of the Dirac sea. Several recent experiments \cite{Hua07,Wil07,Ozy07} have investigated this unusual coupling of electron-like and hole-like dynamics.

\begin{figure}[tb]
\includegraphics[width=1\linewidth]{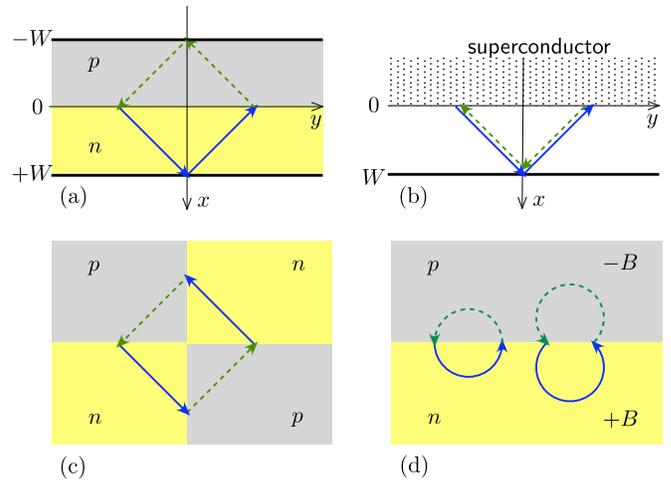}
\caption{\label{NSpn_trajectories}
Periodic orbits in an \textit{n-p} junction (panel a) and in a normal-superconductor (NS) junction (panel b), at $\varepsilon=0$ in the case of an abrupt interface. (Solid and dashed lines distinguish electron-like and hole-like trajectories.) Negative refraction in the \textit{n-p} junction maps onto Andreev retroreflection in the NS junction upon mirroring in the interface at $x=0$. Destructive interference of the electron-like and hole-like segments of the periodic orbit suppresses the density of states at the Fermi level. Panels c and d show alternative geometries that exhibit a suppression of the local density of states in an unbounded system.
}
\end{figure}

In the course of an analysis of these experiments a curious similarity was noticed \cite{Two07} between negative refraction \cite{Che07,Not00} at an \textit{n-p} junction and Andreev retroreflection \cite{And64} at the interface between a normal metal (N) and a superconductor (S). As illustrated in Fig.\ \ref{NSpn_trajectories}a,b, the trajectories at an \textit{n-p} junction and at an NS junction are related by mirroring $x\mapsto -x$ at the interface (taken at $x=0$). Here we show that the similarity is not limited to classical trajectories, but extends to the fully quantum mechanical wave functions and energy spectra. This implies that quantum effects associated with superconductivity, such as the proximity effect and the Josephson effect, have analogues in an \textit{n-p} junction.

We have found a precise mapping between the Dirac Hamiltonian \cite{DiV84} of an \textit{n-p} junction and the Dirac-Bogoliubov-De Gennes Hamiltonian \cite{Bee06} of an NS junction under the condition that the electrostatic potential $U$ in the \textit{n-p} junction is antisymmetric, $U(-x,y)=-U(x,y)$, with respect to the interface. The Fermi level is chosen at zero energy, symmetrically between the \textit{n} and \textit{p} regions. Such a symmetric \textit{n-p} junction turns out to have the same excitation spectrum as an NS junction for excitation energies $\varepsilon$ small compared to the superconducting gap $\Delta$. After presenting the mapping in its mathematical form, we consider the two major physical implications: Zero density of states at the Fermi level and persistent current flow in equilibrium. A comparison with computer simulations of a tight-binding model of graphene is presented at the end of the paper.

{\em Derivation of the mapping.} ---
The correspondence between Klein tunneling and Andreev reflection consists of a mapping of an eigenstate $\Psi$ of the Dirac Hamiltonian $H$ of a symmetric \textit{n-p} junction onto electron and hole eigenstates $\Psi_{e}$ and $\Psi_{h}$ in the normal part $x>0$ of the NS junction. The Dirac Hamiltonian is given (in the valley-isotropic representation) by
\begin{equation}
H=v[(\bm{p}+e\bm{A})\cdot\bm{\sigma}]\otimes\tau_{0}+U\sigma_{0}\otimes\tau_{0},\label{Diractwovalley}
\end{equation}
with $\bm{p}=-i\hbar(\partial/\partial x,\partial/\partial y)$ the momentum operator in the $x-y$ plane of the graphene layer, $\bm{A}=Bx\bm{\hat{y}}$ the vector potential of a perpendicular magnetic field $B$, and $v$ the electron velocity. The Pauli matrices $\sigma_{i}$ and $\tau_{i}$ act, respectively, on the sublattice and valley degree of freedom (with $\sigma_{0}$ and $\tau_{0}$ a $2\times 2$ unit matrix). We introduce the time-reversal operator ${\cal T}=-(\sigma_{y}\otimes\tau_{y}){\cal C}$, with ${\cal C}$ the operator of complex conjugation, and the parity operator ${\cal P}=i(\sigma_{x}\otimes\tau_{0}){\cal R}$, with ${\cal R}$ the operator of reflection ($x\mapsto -x$). The key property of the Dirac Hamiltonian that we need, in order to map the symmetric \textit{n-p} junction onto an NS junction, is the anticommutation relation
\begin{equation}
{\cal TP}H=-H{\cal TP},\label{anticommute}
\end{equation}
satisfied for any $B$ when $U(-x,y)=-U(x,y)$.

Starting from a solution $H\Psi=\varepsilon\Psi$ of the Dirac equation in the \textit{n-p} junction we now construct an eigenstate in the NS junction at the same eigenvalue $\varepsilon$ by means of the transformation
\begin{equation}
\Psi_{e}(x,y)=\Psi(x,y),\;\;
\Psi_{h}(x,y)={\cal P}\Psi(x,y).
\label{PsiehPsirelation}
\end{equation}
According to Refs.\ \cite{Bee06,Tit06} the electron and hole wave functions $\Psi_{e},\Psi_{h}$ in the normal part of the NS junction should satisfy
\begin{equation}
H\Psi_{e}=\varepsilon\Psi_{e},\;\;-{\cal T}H{\cal T}\Psi_{h}=\varepsilon\Psi_{h},\;\;x>0,\label{Psiehequation}
\end{equation}
with a boundary condition at the NS interface that for $|\varepsilon|\ll \Delta$ takes the form
\begin{equation}
\Psi_{h}(0,y)=i(\sigma_{x}\otimes\tau_{0})\Psi_{e}(0,y)\equiv{\cal P}\Psi_{e}(0,y).\label{Psiehrelation}
\end{equation}
The proof of the mapping now follows by inspection: Firstly, Eq.\ \eqref{Psiehequation} results directly from the transformation \eqref{PsiehPsirelation} with the anticommutation relation \eqref{anticommute}. Secondly, since $\Psi$ is continuous at $x=0$, the boundary condition \eqref{Psiehrelation} is automatically satisfied.

The applicability of the mapping extends to the crystallographic edges of the graphene layer in the following way: The edges of the \textit{n-p} junction are described by the boundary condition $\Psi(\bm{r})=M(\bm{r})\Psi(\bm{r})$ for $\bm{r}$ at the edge \cite{McC04,Akh07}. The mapping to an NS junction still holds, provided that $M$ commutes with ${\cal P}$, which requires
\begin{equation}
(\sigma_{x}\otimes\tau_{0})M(x,y)=M(-x,y)(\sigma_{x}\otimes\tau_{0}). \label{MPrelation}
\end{equation}
For example, an armchair edge parallel to the $x$-axis (with $M\propto\sigma_{x}$ independent of $x$) satisfies the requirement \eqref{MPrelation}, but a zigzag edge parallel to the $x$-axis ($M\propto\sigma_{z}$) does not. A pair of zigzag edges at $x=\pm W$ [with $M(\pm W,y)=\pm\sigma_{z}\otimes\tau_{z}$], on the other hand, do satisfy the requirement \eqref{MPrelation}. An infinite mass boundary condition [with $M(\pm W,y)=\pm\sigma_{y}\otimes\tau_{z}$], likewise, satisfies this requirement.

{\em Suppression of the density of states.} ---
We have calculated the density of states $\rho(\varepsilon)$ by solving the Dirac equation in the \textit{n-p} junction of Fig.\ \ref{NSpn_trajectories}a. The Fermi level (taken at $\varepsilon=0$) is separated from the Dirac point by the energy $\pm\mu$ in the \textit{n} and \textit{p} regions. We take an abrupt interface (width $d$ small compared to the Fermi wave length $\lambda_{F}=hv/\mu$) and wide and long \textit{n} and \textit{p} regions (width $W\gg \lambda_{F}$, length $L\gg W$). The precise choice of boundary condition at $x=\pm W$ does not matter in this regime, as long as it  preserves the symmetry of  the geometry. 

\begin{figure}[tb]
\includegraphics[width=.9\linewidth]{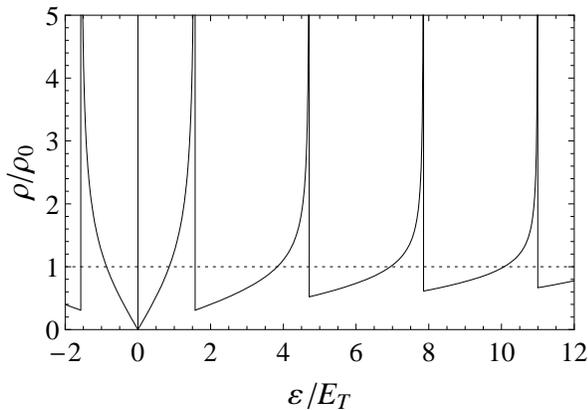}
\caption{\label{rhopn}
Density of states in the \textit{n-p} junction of Fig.\ \ref{NSpn_trajectories}a, calculated from Eq.\ \eqref{dispersion}. The dotted line is the value in the isolated \textit{n} and \textit{p} regions, which is energy independent for $|\varepsilon|\ll\mu$. The density of states vanishes at the Fermi level ($\varepsilon=0$), according to Eq.\ \eqref{rhoresult}.
}
\end{figure}

The calculation for the bipolar junction follows step-by-step the analogous calculation for the Josephson junction in Ref.\ \cite{Tit07}. The dispersion relation (smoothed over rapid oscillations) is given by
\begin{equation}
\varepsilon_{m}(q)=\pi E_{T}(m+\tfrac{1}{2})\sqrt{1-(\hbar vq/\mu)^{2}},\;\;|\varepsilon|\ll\mu,\label{dispersion}
\end{equation}
with $m=0,\pm 1,\pm 2,\ldots$ the mode index and $\hbar q$ the momentum parallel to the \textit{n-p} interface. (The energy $E_{T}=\hbar v/2W$ is the Thouless energy, which is $\ll\mu$ for $W\gg\lambda_{F}$.) The resulting density of states $\rho(\varepsilon)=(4/\pi)\sum_{m}|\partial\varepsilon_{m}/\partial q|^{-1}$ is plotted in Fig.\ \ref{rhopn}. It vanishes linearly as
\begin{equation}
\rho(\varepsilon)=\rho_{0}|\varepsilon|/E_{T}\label{rhoresult}
\end{equation}
for small $|\varepsilon|$, with $\rho_{0}=(2\mu/\pi)(\hbar v)^{-2}$ the density of states (per unit area and including spin plus valley degeneracies) in the separate \textit{n} and \textit{p} regions. This suppression of the density of states at the Fermi level by a factor $\varepsilon/E_{T}$ is precisely analogous to an NS junction, where the density of states is suppressed by the superconducting proximity effect (compare, for example, our Fig.\ \ref{rhopn} with Fig.\ 8 of Ref.\ \cite{Tit07}). In particular, the peaks in $\rho(\varepsilon)$ at $\varepsilon=\pi E_{T}(m+\tfrac{1}{2})$ are analogous to the De Gennes-Saint James resonances in Josephson junctions \cite{DeG63}.

In a semiclassical description, the suppression of the density of states in the \textit{n-p} junction can be understood as destructive interference of the electron-like and hole-like segments of a periodic orbit (solid and dashed lines in Fig.\ \ref{NSpn_trajectories}a). At the Fermi level, the dynamical phase shift accumulated in the \textit{n} and \textit{p} regions cancels, and what remains is a Berry phase shift of $\pi$ from the rotation of the pseudospin of a Dirac fermion \cite{Nov05,Zha05}.

\begin{figure}[tb]
\includegraphics[width=.9\linewidth]{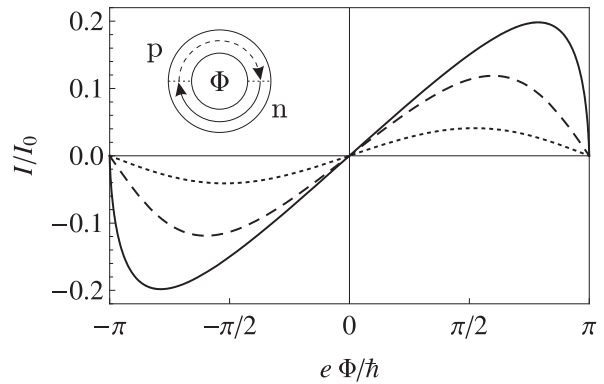}
\caption{\label{current}
Persistent current through a ring containing an abrupt \textit{n-p} interface, as a function of the magnetic flux through the ring. The solid curve is for zero temperature $T=0$, the dashed curve for $T=E_{T}/4k_{B}$, and the dotted curve for $T=E_{T}/2k_{B}$.
}
\end{figure}

{\em Persistent current.} ---
If the \textit{n} and \textit{p} regions enclose a magnetic flux $\Phi$, as in the ring geometry of Fig.\ \ref{current} (inset), then the Berry phase shift can be compensated and the suppression of the density of states can be eliminated. The resulting flux dependence of the ground state energy $E={\cal A}\int_{-\infty}^{0}\rho(\varepsilon)\varepsilon\,d\varepsilon$ (with ${\cal A}$ the joint area of the \textit{n} and \textit{p} regions) implies that a current $I=dE/d\Phi$ will flow through the ring in equilibrium at zero temperature, as in a Josephson junction \cite{Imr97}. According to Eq.\ \eqref{rhoresult}, the order of magnitude
\begin{equation}
I_{0}=(e/\hbar)E_{T}^{2}/\delta=(e/\hbar)NE_{T}\label{I0def}
\end{equation}
of this persistent current is set by the level spacing $\delta=({\cal A}\rho_{0})^{-1}$ and by the Thouless energy $E_{T}=\hbar v/\pi r=N\delta$ in the ring geometry (of radius $r$ and width $w\ll r$, supporting $N=4\mu w/\pi\hbar v\gg 1$ propagating modes). Because of the macroscopic suppression of the density of states, this is a macroscopic current --- larger by a factor $N$ than the mesoscopic persistent current in a ballistic metal ring \cite{Imr97,But83}.

We have calculated $I(\Phi)$ for a simple model of an abrupt \textit{n-p} junction in an $N$-mode ring without intermode scattering, neglecting the effect of the curvature of the ring on the spectrum and also assuming that the magnetic field is confined to the interior of the ring. (These approximations are reasonable for $\lambda_{F}\ll w\ll r$.) The slowly converging, oscillatory integral over $\rho(\varepsilon)$ was converted into a rapidly decaying sum over Matsubara frequencies by the method of Ref.\ \cite{Bro97}. The zero-temperature result is plotted in Fig.\ \ref{current} (solid curve). The maximal persistent current is $I_{c}\approx 0.2\,I_{0}$. This is the same value, up to a numerical coefficient, as the critical current of a ballistic Josephson junction \cite{note1}. We have also included the results at finite temperature, showing the decay when the thermal energy $k_{B}T\simeq E_{T}$.  

\begin{figure}[tb]
\includegraphics[width=.9\linewidth]{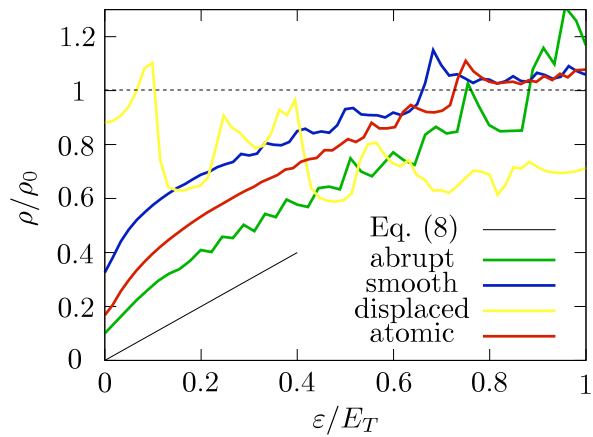}
\caption{\label{simulation}
Same as Fig.\ \ref{rhopn}, but now calculated from a tight-binding model of graphene (lattice constant $a$, $W/a=400$). The colors distinguish different values of $\lambda_{F}$ and $d$, corresponding to an abrupt interface ($\lambda_{F}/a=65$, $d/a=12$), a smooth interface ($\lambda_{F}/a=12$, $d/a=12$), and an atomically sharp interface ($\lambda_{F}/a=12$, $d/a\simeq 1$). The suppression of the density of states vanishes if the reflection symmetry is broken by displacing the interface (yellow curve, $\lambda_{F}/a=65$, $d/a=12$, displacement $=65\,a$).
}
\end{figure}

{\em Comparison with computer simulations.} --- 
To test our analytical predictions, we have performed computer simulations of a tight-binding Hamiltonian on a honeycomb lattice (lattice constant $a$). We took a symmetric \textit{n-p} junction with zigzag boundaries at $x=\pm W$ (with $W/a=400$) and calculated the density of states $\rho(\varepsilon)$, smoothed by a Lorentzian (width $0.01\,E_{T}$) to eliminate the rapid oscillations. Results are shown in Fig.\ \ref{simulation}, for different Fermi wave lengths $\lambda_{F}=hv/\mu$ and widths $d$ of the \textit{n-p} interface [potential profile $U(x)=-\mu \tanh(4x/d)$]. A clear suppression of $\rho(\varepsilon)$ is observed within an energy range $E_{T}$ from the Fermi level at $\varepsilon=0$. The suppression is somewhat smaller than predicted by Eq.\ \eqref{rhoresult} (black solid line), in particular for $d\simeq a$ (red curve, when the Dirac equation no longer applies) and for $d\gtrsim\lambda_{F}$ (blue curve, when Klein tunneling happens only near normal incidence \cite{Che06}).

As expected, the suppression is sensitive to perturbations of the reflection symmetry. For example, as shown in Fig.\ \ref{simulation} (yellow curve), a displacement of the \textit{n-p} interface by $\lambda_{F}$ spoils the systematic destructive interference due to the Berry phase, and thus eliminates the suppression of the {\em global\/} density of states.

We would still expect an effect on the {\em local\/} density of states if we could confine the carriers to the \textit{n-p} interface. This might be achieved by means of the saddle point potential $U=\mu\,{\rm sign}\,(xy)$ of Fig.\ \ref{NSpn_trajectories}c, or by means of the nonuniform magnetic field $B=B_{0}x$ of Fig.\ \ref{NSpn_trajectories}d. Destructive interference of the periodic orbits in each of these unbounded geometries will suppress the local density of states near the interface by the same mechanism as in the confined geometry of Fig.\ \ref{NSpn_trajectories}a. Because of disorder, the suppression will be limited to a mean free path or corrugation length from the \textit{n-p} interface. Since the predicted suppression of the density of states at the Fermi level happens at a large energy separation $\mu$ from the Dirac point (see Fig.\ \ref{Klein_tunneling}), it should be distinguishable in a local measurement (for example, by a tunneling probe) from any features associated with the conical singularity in the band structure at the Dirac point.

From a different perspective, the correspondence derived here offers the intriguing opportunity to observe superconducting analogies in non-electronic systems governed by the same Dirac equation as graphene. An example would be a two-dimensional photonic crystal on a honeycomb or triangular lattice \cite{Rag06,Sep07}, in which the analogue of an \textit{n-p} junction has been proposed recently \cite{Gar07}. The detrimental effects of disorder should be relatively easy to avoid in such a metamaterial.

We acknowledge discussions with J. Nilsson and R. A. Sepkhanov. This research was supported by the Dutch Science Foundation NWO/FOM.

\appendix
\section{Calculation of the persistent current and comparison with supercurrent}

In this Appendix we present the calculation leading to the persistent current through the bipolar junction plotted in Fig.\ \ref{current}. We follow closely the analogous calculation for the supercurrent through a Josephson junction of Ref.\ \cite{Bro97}, and compare the two systems at the end. For the sake of this comparison, it is convenient to work with the density of states $\tilde{\rho}=({\cal A}/2)\rho$ per spin direction, integrated over the area ${\cal A}$ of the system. We will likewise, in this Appendix, count the number of propagating modes $\tilde{N}=N/2$ per spin direction.

\subsection{Persistent current}

The persistent current $I=dF/d\Phi$ at temperature $T$ is given by the derivative of the free energy $F$ with respect to the flux $\Phi$ enclosed by the ring containing the \textit{n-p} junction. This can be expressed as an integral over the density of states,
\begin{equation}
I=-2k_{B}T\frac{d}{d\Phi}\int_{-\infty}^{\infty}d\varepsilon\,\tilde{\rho}(\varepsilon)\ln[2\cosh(\varepsilon/2k_{B}T)].\label{Fdef}
\end{equation}
We have set the Fermi energy at zero and used the electron-hole symmetry $\tilde{\rho}(\varepsilon)=\tilde{\rho}(-\varepsilon)$. The factor of two in front accounts for the two spin directions (which are not counted separately in $\tilde{\rho}$).

Since the spectrum of the ring is discrete, the density of states $\tilde{\rho}(\varepsilon)=\sum_{i}\delta(\varepsilon-\varepsilon_{i})$ consists of delta functions at the solutions of the equation
\begin{equation}
{\cal F}(\varepsilon)\equiv{\cal F}_{0}(\varepsilon)\prod_{i}(\varepsilon-\varepsilon_{i})=0.\label{Fequation}
\end{equation}
(The  index $i$ counts spin-degenerate levels once.) The function ${\cal F}_{0}$ is $>0$ and even in $\varepsilon$, but can otherwise be freely chosen. The density of states is then written as
\begin{equation}
\tilde{\rho}(\varepsilon)=-\frac{1}{\pi}\frac{d}{d\varepsilon}\,{\rm Im}\,\ln{\cal F}(\varepsilon+i0^{+}),\label{rhoFrelation}
\end{equation}
with $0^{+}$ a positive infinitesimal.

Substitution of Eq.\ \eqref{rhoFrelation} into Eq.\ \eqref{Fdef} gives, using again the electron-hole symmetry,
\begin{equation}
I=\frac{2k_{B}T}{\pi i}\frac{d}{d\Phi}\int_{-\infty+i0^{+}}^{\infty+i0^{+}}d\varepsilon\,\ln[2\cosh(\varepsilon/2k_{B}T)]\frac{d}{d\varepsilon}\ln{\cal F}(\varepsilon).\label{Fintegral}
\end{equation}
The expression for the persistent current becomes, upon partial integration,
\begin{equation}
I=-\frac{1}{\pi i}\frac{d}{d\Phi}\int_{-\infty+i0^{+}}^{\infty+i0^{+}}d\varepsilon\,\tanh(\varepsilon/2k_{B}T)\ln{\cal F}(\varepsilon).\label{Iintegral}
\end{equation}
We close the contour in the upper half of the complex plane. We assume that ${\cal F}_{0}$ is chosen such that $\ln{\cal F}$ has no singularities for ${\rm Im}\,\varepsilon>0$. The only poles of the integrand in Eq.\ \eqref{Iintegral} then come from the hyperbolic tangent, at the Matsubara frequencies $i\omega_{n}=(2n+1)i\pi k_{B}T$. Summing over the residues we arrive at the expression \cite{Bro97}
\begin{equation}
I=-4k_{B}T\frac{d}{d\Phi}\sum_{n=0}^{\infty}\ln{\cal F}(i\omega_{n}).\label{Isum}
\end{equation}

In our model of an $\tilde{N}$-mode ring without intermode scattering we can calculate separately the contribution to $I$ from each propagating mode, with transverse momentum $q_{m}$. The total current is then a sum over these contributions,
\begin{equation}
I=-4k_{B}T\sum_{m=1}^{\tilde{N}}\frac{d}{d\Phi}\sum_{n=0}^{\infty}\ln{\cal F}(i\omega_{n},q_{m}).\label{Isum2}
\end{equation}
The function ${\cal F}(\varepsilon,q)$, which determines the energy levels in the bipolar junction for a given transverse mode, is the limit $\Delta\rightarrow\infty$ of the analogous function in a Josephson junction \cite{Tit06}. We find
\begin{align}
{\cal F}(\varepsilon,q)={}&\frac{\mu^{2}-\varepsilon^{2}+(\hbar vq)^{2}}{\theta_{+}\theta_{-}E_{T}^{2}}\sin \theta_{+}\sin \theta_{-}\nonumber\\
&+\cos \theta_{+}\cos \theta_{-}+\cos(e\Phi/\hbar),\label{Fdef1}\\
\theta_{\pm}={}&E_{T}^{-1}\sqrt{(\mu\pm\varepsilon)^{2}-(\hbar vq)^{2}}.\label{Fdef2}
\end{align}
Substitution into Eq.\ \eqref{Isum2} gives the persistent current,
\begin{equation}
I=4k_{B}T\frac{e}{\hbar}\sin(e\Phi/\hbar)\sum_{m=1}^{\tilde{N}}\sum_{n=0}^{\infty}\frac{1}{{\cal F}(i\omega_{n},q_{m})}.\label{Isum3}
\end{equation}

We wish to evaluate the expression \eqref{Isum3} in the regime $\mu\gg E_{T}$, $\tilde{N}\gg 1$. The sum over modes may be replaced by an integral, according to $\sum_{m=1}^{\tilde{N}}\rightarrow (\tilde{N}/k_{F})\int_{0}^{k_{F}}dq$, with $k_{F}=\mu/\hbar v$ the Fermi wave vector. Since the sum over the Matsubara frequencies converges exponentially fast for $\omega_{n}\gtrsim E_{T}$, we can also assume $\mu\gg\omega_{n}$. In this large-$\mu$ regime we may approximate $\theta_{\pm}\approx \alpha\pm i\Omega_{n}$, with 
\begin{align}
\alpha&=(\mu/E_{T})[1-(q/k_{F})^{2}]^{1/2},\label{alphadef}\\
\Omega_{n}&=(\omega_{n}/E_{T})[1-(q/k_{F})^{2}]^{-1/2}.\label{Omegadef}
\end{align}
The function ${\cal F}$ takes the form
\begin{align}
&{\cal F}(i\omega_{n},q)=X\cos^{2}\alpha+Y\sin^{2}\alpha,\label{FXY}\\
&X=Z\sinh^{2}\Omega_{n}+\cosh^{2}\Omega_{n}+\cos(e\Phi/\hbar),\label{Xdef}\\
&Y=Z\cosh^{2}\Omega_{n}+\sinh^{2}\Omega_{n}+\cos(e\Phi/\hbar),\label{Ydef}\\
&Z=\frac{\mu^{2}+(\hbar vq)^{2}+\omega_{n}^{2}}{\mu^{2}-(\hbar vq)^{2}+(E_{T}\Omega_{n})^{2}}\approx\frac{1+(q/k_{F})^{2}}{1-(q/k_{F})^{2}}.\label{Zdef}
\end{align}

The phase $\alpha$ varies rapidly as a function of $q$, so we average $1/{\cal F}$ first over this phase,
\begin{equation}
\frac{1}{{\cal F}}\rightarrow\int_{0}^{2\pi}\frac{d\alpha}{2\pi}\frac{1}{X\cos^{2}\alpha+Y\sin^{2}\alpha}=\sqrt{\frac{1}{XY}}.\label{Faverage}
\end{equation}
We substitute Eq.\ \eqref{Faverage} into Eq.\ \eqref{Isum3} and evaluate it numerically, to arrive at the curves of $I$ versus $\Phi$ shown in Fig.\ \ref{current}.

\subsection{Comparison with supercurrent}

\begin{figure}[tb]
\includegraphics[width=.9\linewidth]{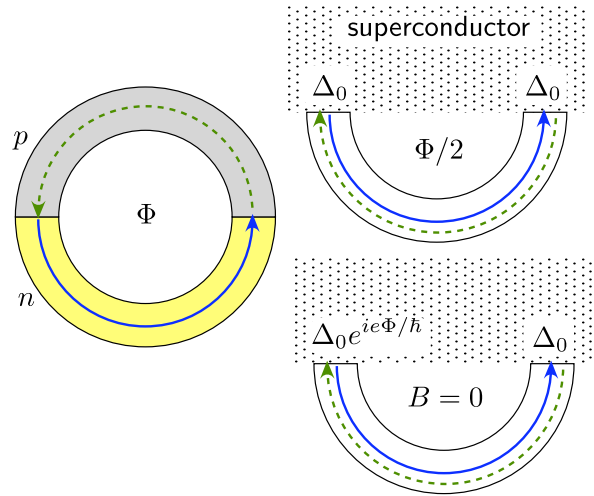}
\caption{\label{ring}
Mapping of a bipolar ring containing two \textit{n-p} junctions (left panel) onto a Josephson ring containing two NS junctions (right panels), by mirroring the hole-like trajectories (dashed) in the line through the interfaces. The persistent current $I$ through the bipolar ring at the left maps onto a supercurrent $I_{J}$ through a Josephson ring at the right. Because the enclosed flux $\Phi$ is halved by the mapping, the $h/e$-periodicity of $I$ maps onto an $h/2e$ periodicity of $I_{J}$. The flux enclosed by the Josephson ring in the upper right panel may be gauged away, with the introduction of a phase difference $\phi=e\Phi/\hbar$ between the order parameters at the two NS interfaces (lower right panel).
}
\end{figure}

The mapping between bipolar junctions and Josephson junctions is illustrated in Fig.\ \ref{ring}. Instead of a ring geometry we may equivalently consider a planar SNS junction, with a phase difference $\phi$ between the two superconducting reservoirs. In the absence of mode mixing the two geometries carry the same supercurrent $I_{J}$, at the same number of transverse modes $\tilde{N}$. (The Thouless energy in the SNS junction is $E_{T}=\hbar v/L$, with $L$ the separation of the two NS interfaces.)

The supercurrent $I_{J}$ through the Josephson junction is given by \cite{Bee92}
\begin{equation}
I_{J}=-2k_{B}T\frac{2e}{\hbar}\frac{d}{d\phi}\int_{0}^{\infty}d\varepsilon\,\rho_{J}(\varepsilon)\ln[2\cosh(\varepsilon/2k_{B}T)],\label{FJdef}
\end{equation}
with $\phi$ the phase difference across the junction and $\rho_{J}$ the density of states per spin direction. The mapping relates $\phi\leftrightarrow e\Phi/\hbar$ (see Fig.\ \ref{ring}) and $\rho_{J}\leftrightarrow\tilde{\rho}$. Comparison of Eqs.\ \eqref{Fdef} and \eqref{FJdef} then shows that $I_{J}(\phi)\leftrightarrow I(\Phi)$. The bipolar junction and Josephson junction therefore carry the same current in equilibrium.

The result in the literature \cite{Ish70,Bar72,Svi73,But86} for a ballistic SNS junction is a piecewise linear dependence of $I_{J}$ on $\phi$ at zero temperature, close to but not identical to the solid curve in Fig.\ \ref{current}. As we will now show, the difference is due to the presence or absence of a step in the Fermi energy at the NS interfaces.

\begin{figure}[tb]
\includegraphics[width=.9\linewidth]{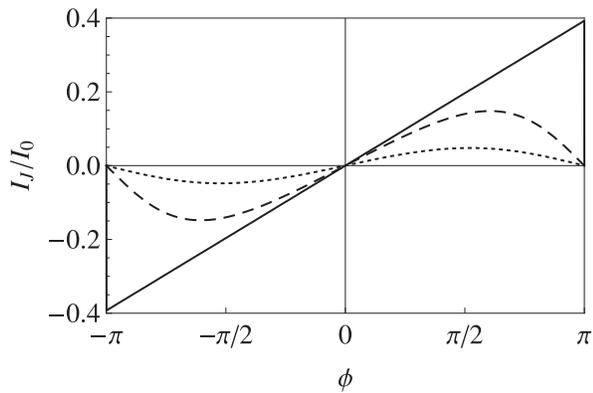}
\caption{\label{currentJ}
Supercurrent through a ballistic Josephson junction as a function of the phase difference $\phi$ between the two superconducting reservoirs, calculated from Eq.\ \eqref{IJ1}. The solid curve is for zero temperature $T=0$, the dashed curve for $T=E_{T}/4k_{B}$, and the dotted curve for $T=E_{T}/2k_{B}$. The difference with the analogous result for a bipolar junction in Fig.\ \ref{current} arises because this figure is for equal Fermi energy $\mu_{S}=\mu$ in superconductor and normal metal, while Fig.\ \ref{current} maps onto a Josephson junction with $\mu_{S}\gg\mu$.  
}
\end{figure}

On the one hand, the mapping between bipolar and Josephson junctions relies on the boundary condition \eqref{Psiehrelation} at the NS interface, which assumes that the Fermi energy $\mu_{S}$ in the superconductor is much larger than the value $\mu$ in the normal region \cite{Tit06}. On the other hand, Refs.\ \cite{Ish70,Bar72,Svi73,But86} assume $\mu_{S}=\mu$. The function ${\cal F}(\varepsilon,q)$ is then given by
\begin{equation}
{\cal F}(\varepsilon,q)=\cos(\theta_{+}-\theta_{-})+\cos\phi,\label{FSNS}
\end{equation}
resulting in
\begin{equation}
I_{J}(\phi)=4k_{B}T\frac{e}{\hbar}\sum_{m=1}^{\tilde{N}}\sum_{n=0}^{\infty}\,\frac{\sin\phi}{\cosh 2\Omega_{n}(q_{m})+\cos\phi}.\label{IJ1}
\end{equation}
The resulting supercurrent is plotted in Fig.\ \ref{currentJ}, for different temperatures.

At $T=0$ the sum over $n$ reduces to an integral, $\sum_{n=0}^{\infty}\rightarrow(2\pi k_{B}T)^{-1}\int_{0}^{\infty}d\omega$, which evaluates to
\begin{align}
I_{J}(\phi)={}&\frac{2eE_{T}}{\pi\hbar}\sum_{m=1}^{\tilde{N}}[1-(q_{m}/k_{F})^{2}]^{1/2}\nonumber\\
&\times\int_{0}^{\infty}d\omega\,\frac{\sin\phi}{\cosh 2\omega+\cos\phi}\nonumber\\
={}&\phi\,\frac{eE_{T}}{\pi\hbar}\sum_{m=1}^{\tilde{N}}[1-(q_{m}/k_{F})^{2}]^{1/2}, \;\;|\phi|<\pi.\label{IJ2}
\end{align}
(The $\phi$-dependence is repeated periodically outside of the interval $-\pi<\phi<\pi$.) We thus recover the piecewise linear $\phi$-dependence of the supercurrent \cite{Ish70,Bar72,Svi73,But86}.

For $\tilde{N}\gg 1$ the sum over modes may also be evaluated as an integral, $\sum_{m=1}^{\tilde{N}}\rightarrow(\tilde{N}/k_{F})\int_{0}^{k_{F}}dq$, with the result
\begin{equation}
I_{J}(\phi)=\phi\,\frac{e\tilde{N}E_{T}}{4\hbar},\;\;|\phi|<\pi.\label{IJ3}
\end{equation}
The critical current $I_{c}=\pi e\tilde{N}E_{T}/4\hbar=(\pi/8)I_{0}$ is about two times larger than the maximal persistent current $I_{c}\approx 0.2\,I_{0}$ found in the bipolar junction, because of the absence of a step in the Fermi energy at the NS interfaces.

Eq.\ \eqref{IJ3} holds in a two-dimensional geometry. In three dimensions the sum over modes becomes $\sum_{m=1}^{\tilde{N}}\rightarrow(2\tilde{N}/k_{F}^{2})\int_{0}^{k_{F}}qdq$, resulting in
\begin{equation}
I_{J}(\phi)=\phi\,\frac{2e\tilde{N}E_{T}}{3\pi\hbar},\;\;|\phi|<\pi,\label{IJ4}
\end{equation}
in agreement with Refs.\ \cite{Bar72,Svi73}. (The numerical coefficient in Ref.\ \cite{Ish70} is different.) In the one-dimensional case $\tilde{N}=1$ of a single spin-degenerate mode (group velocity $v_{\rm group}=v[1-(q_{1}/k_{F})^{2}]^{1/2}$) we find instead
\begin{equation}
I_{J}(\phi)=\phi \frac{ev_{\rm group}}{\pi L},\;\;|\phi|<\pi,\label{IJ5}
\end{equation}
in agreement with Ref.\ \cite{But86} (up to a factor of two, presumably because Ref.\ \cite{But86} does not account for the spin degeneracy of the mode).


\begin{thebibliography}{99}
\bibitem{note1} The finite-temperature calculation of $I(\Phi)$ is worked out in the Appendix, and compared with the analogous result in a Josephson junction.
\end{thebibliography}
\end{document}